\documentclass[12pt]{article}

\usepackage{latexsym}
\usepackage{amsmath,amssymb}
\usepackage[hang,flushmargin]{footmisc}

  \def\d{\delta} \def\D{\Delta}
    \def\g{\gamma}
\def\G{\Gamma}  

   \def\Th{\Theta}  \def\l{\lambda}
   \def\p{\pi}
\def\r{\rho}
  
\def\t{\tau}   \def\Om{\Omega}

\def\cW{{\cal W}}

\newcommand{\ooi}{(1+o(1))}

\newcommand{\ul}[1]{\mbox{\boldmath$#1$}}
\newcommand{\sul}[1]{\mbox{\scriptsize\boldmath$#1$}}

\newcommand{\wh}[1]{\widehat{#1}}

\newcommand{\rdown}[1]{{\left\lfloor #1\right \rfloor}}

\newcommand{\brac}[1]{\left(#1\right)}

\newcommand{\bfrac}[2]{\left(\frac{#1}{#2}\right)}

\newcommand{\rai}{\rightarrow \infty}

\def\E{\mathbf{E}}

\def\Pr{\mbox{{\bf Pr}}}
\def\whp{{\bf whp}}

\newcommand{\beq}[1]{\begin{equation}\label{#1}}
\newcommand{\eeq}{\end{equation}}

\newtheorem{theorem}{Theorem}
\newtheorem{lemma}{Lemma}

\newcommand{\ignore}[1]{}

\newcommand{\proofstart}{{\noindent \bf Proof. }}

\newcommand{\proofend}{\hspace*{\fill}\mbox{$\Box$}}

\begin{document}

\title{Coalescing Random Walks and Voting on Connected Graphs\thanks{Partially supported by
the Royal Society International Joint Project grant JP090592
''Random Walks, Interacting Particles and Faster Network Exploration,''
and the EPSRC grant EP/J006300/1
''Random walks on computer networks."
A Preliminary version of the results in this paper was presented in the Proceedings of PODC 2012~\cite{CEOR-PODC2012}.
}}

\author{
Colin Cooper\thanks{Department of Informatics, King's College London, UK}
\and Robert Els\"asser\thanks{Department of Computer Science, University of Salzburg, Austria}
\and Hirotaka Ono\thanks{Department of Economic Engineering, University of Kyushu, Fukuoka, Japan}
\and Tomasz Radzik\thanks{Department of Informatics, King's College London, UK}
}

\date{\today}

\maketitle
\makeatother

\begin{abstract}
In a {\em coalescing random walk}, a set of particles make
independent
discrete-time
random
walks on a graph.
Whenever one or more particles meet at a vertex,  they unite to form
a single particle,
which then continues a random walk through the graph.

Let $G=(V,E)$, be an undirected and connected  graph, with $n$ vertices and $m$ edges.
The {\em coalescence time}, $C(n)$, is the expected time for
all particles to coalesce, when initially one particle is located at each vertex.
We study the problem of bounding the coalescence time
for general connected graphs, and  prove that
\[
C(n) = O\brac{\frac{1}{1-\l_2}\brac{\log^{4} n +  \frac{n}{\nu}}}.
\]
Here $\l_2$ is the second eigenvalue of the transition  matrix of the random walk.
To avoid problems arising from e.g.
lack of coalescence on bipartite graphs, we assume the random walk can be made lazy if
required.
The value of $\nu$ is given by
$\nu= \sum_{v\in V} d^2(v)/(d^2n)$, where
$d(v)$ is the degree of vertex $v$, and
$d=2m/n$ is the average degree.
The parameter $\nu$ is an indicator of the variability of vertex degrees: $1
\le \nu = O(n)$, with $\nu=1$ for regular graphs.

Our general bound on $C(n)$ holds
for all connected graphs.
This implies, for example,
that $C(n)=O(n/(1-\l_2))$ for $d$-regular graphs with expansion parameterized by the eigenvalue gap
$1-\l_2$.
The  bound on $C(n)$ given above
is sub-linear for some classes of graphs with skewed degree distributions.

In the  {\em voter model},
initially each vertex has a distinct opinion,
and at each step each vertex changes its opinion to that of a random neighbour.
Let $\E (C_{\ul v})$ be the expected time  for voting to complete, that is, for a unique opinion to emerge.
A system of coalescing particles, where initially one particle
is  located at each vertex,  corresponds to the voter model in that $\E(C_{\ul v})=C(n)$.
Thus our result stated  above  for $C(n)$ also gives general bounds for
$\E(C_{\ul v})$.
\end{abstract}

\section{Introduction}

In a {\em coalescing random walk}, a set of particles make
independent discrete-time random
walks in an undirected connected graph.
Whenever two or more particles meet at a vertex, then they unite to form a single particle
which then continues to make a random walk through the graph.

Let  $G=(V,E)$ be an undirected connected graph with $n$ vertices and $m$ edges.
The {\em coalescence time} is the expected time for
all particles to coalesce, when initially one particle is located at each vertex of the graph.
We study the problem of bounding the coalescence time
for general connected graphs.

For a given graph $G$ we denote the coalescence time of an $n$ particle system by  $C(n)=C_G(n)$.
In order to bound $C(n)$, we study the coalescence time $C(k)=C_G(k)$ of a system of $k$ particles
for any $2 \le k \le n$. The expected time for the $k$ particles to coalesce to a single particle
depends on their initial positions.
Let $C_k(i_1,...,i_k)$, be the coalescence time when the
 particles start from distinct vertices $i_1,...,i_k$.
The worst case expected coalescence time for $k$ particles is
\[ C(k)= \max_{i_1,...,i_k} \E (C_k(i_1,...,i_k)).\]
In the special case of two particles,
$C(2)$ is more naturally referred to as
the (worst case expected) {\em meeting time\/} of two random walks.

A system of $n$ coalescing particles where initially one particle
is  located at each vertex,  corresponds to another classical
problem, the {\em voter model}, which is
defined as follows.
Initially each vertex has a distinct opinion,
and at each step each vertex changes its opinion to that of a random neighbour.

Let $C_{\ul v}$ be the number of steps
for voting to be completed, i.e., for a unique opinion to emerge.
The expected completion time of voting, $\E (C_{\ul v})$, is  called the {\em voting time}.
The random variable $C_{\ul v}$
has the same distribution, and hence the same expected value,
as the coalescence time $C_n$ of $n$ coalescing particles, where one particle
is initially located at each vertex (see~\cite{AlFi}).
Thus $C(n) \equiv \E(C_n) = \E(C_{\ul v})$, and any bound on coalescence time $C(n)$
applies equally to the voting time  $E(C_{\ul v})$. As the coalescence time is easier to estimate,
we focus on this quantity henceforth.

The coalescing random walk is the key ingredient in the self-stabilizing
mutual exclusion algorithm of Israeli and Jalfon \cite{IJ}.
Initially each vertex emits a token which makes a random walk on $G$.
On meeting at a vertex, tokens coalesce.
Provided the graph is connected, and not bipartite,  eventually  only one token will remain,
and the vertex with the token has exclusive access to some
resource. The token makes a random walk on $G$, so in the long run it  will visit all vertices of $G$
in proportion to their stationary distribution.

\subsection*{Previous work on coalescing random walks}

We  summarize some known results for coalescing random walks.
There are two distinct models for the transition times of random walks on finite graphs.
In the  {\em discrete-time} model, all
walks make transitions synchronously at steps $t=1,2,..$.
In the {\em continuous-time\/} model, each walk $W$
waits for a random time $t_W$ independently of other walks, and then makes a transition.
The wait time $t_W$ is an independent exponential random variable with rate 1.

Let $H_{u,v}$ denote the {\em hitting time\/} of vertex $v$ starting from vertex $u$, that is,
the random variable which gives the time taken for a random walk
starting from vertex $u$ to reach vertex $v$;
and let $H_{\max} = \max_{u,v}  \E(H_{u,v})$.
Aldous~\cite{Aldous-MeetingTimes} considers
$C(2)$, the meeting time of two random walks,
in the continuous-time model, and shows  that
\[
C(2)= \Om(m/\D) \quad \text{and}\quad  C(2)= O(H_{\max}),
\]
where $\D$ is the maximum degree of $G$.
These upper and lower bounds can be far apart,
e.g. for a star graph (with loops),
$C(2)= \Th(1)$ whereas $H_{\max}= \Th(n)$.

The $O(H_{\max})$ bound on $C(2)$ implies  that $C(n) = O(H_{\max}\log n)$,
since the number of particles halves in $O(H_{\max})$ time.
Aldous~\cite{Aldous-MeetingTimes} conjectured that $C(n)$ is actually $O(H_{\max})$.
Earlier results by Cox~\cite{Cox-1989} for the continuous-time model,
imply $C(n) = O(H_{\max})$
for constant dimension tori and grids.

For regular graphs, in the continuous-time model,
Aldous and Fill~\cite{AlFi}
show that,
$C(n)$ $\le$ $e(\log n+2) H_{\max}$,
$C(n)$ $\le$ $rn^2/(4s)$ for $r$-regular $s$-edge connected graphs, and
$C(n) \sim n$ for complete graphs.
Cooper {\em et al.}~\cite{CFR} confirmed that the conjecture $C(n)=O(H_{\max})$
holds for discrete-time random walks on random regular graphs.
This follows from their result that
for $r$-regular random graphs
$C(n) \sim 2 ((r-1)/(r-2))n$,
with high probability.
We use the notation {\em with high
probability} (\whp), to mean  with probability tending to 1 as
 $n \rai$. The notation $f(n) \sim g(n)$ means that $f(n) = (1 \pm o(1))g(n)$.

Simple bounds on $H_{\max}$ can be obtained from
 the commute time between any pair of vertices (see e.g. Corollary 3.3 of Lovasz \cite{Lo}).
For a graph $G$ with $n$ vertices,
$m$ edges
and  minimum degree $\d(G)$, we have
\begin{equation}\label{nciwbc34}
\frac{m}{2\d(G)} \le H_{\max} \le \frac{4m}{(1-\l_2)\d(G)}.
\end{equation}
As $\d(G) \le d$ the average degree, it follows that $H_{\max} \ge n/4$ for any graph. An upper bound,  for connected graphs,
of $H_{\max} \le 4m/(1-\l_2)$ follows from $\d(G) \ge 1$.

\ignore{
Finally, we mention a particular variant of the voter model, the {\em two-party model},
in which there are two initial opinions {\em A} and {\em B}.
Voting is completed when all vertices have the same opinion.
Donnelly and Welsh~\cite{DonnellyWelsh-1983}
considered
the continuous-time two-party voter model and its relation to the continuous-time coalescing random walks.
Hassin and Peleg~\cite{HassinPeleg-InfComp2001}
and Nakata {\em et al.}~\cite{Nakata_etal_1999} considered
the {\em discrete-time\/} two-party voter model, and
discussed its application to consensus  problems in distributed systems.
Both papers~\cite{HassinPeleg-InfComp2001} and~\cite{Nakata_etal_1999} focus on analysing the probability that all vertices
will eventually adopt the opinion which is initially held by a given group of vertices.
The central result is that the probability that opinion {\em A} wins is
$d(${\em A}$)/(2m)$, where
$d(${\em A}$)$ is the sum of the degrees of the vertices initially holding opinion {\em A},
and $m$ is the number of edges in $G$.
}

\subsection*{General results for coalescing walks on graphs}

In this article, we study the problem of bounding the coalescence time $C(n)$
of any connected graph.
We assume that the graphs $G$ we consider are not bipartite, or that if
$G$ is bipartite, then the random walks
are lazy and pause with  probability $1/2$ at each step.
Equivalently, for  the voting process, we assume that vertices may choose their own opinion
with this probability.

Our main result, stated formally below, is given in terms of
the second eigenvalue of the transition  matrix of the random walk, $\l_2$,
and a parameter $\nu$ which measures the variability of the degree sequence.
Let $d(v)$ be the degree of vertex $v$, and $d=2m/n$  the average degree.
The parameter $\nu$, the ratio of the average squared degree to the average degree squared,  is
defined as
\begin{equation}\label{nu-definition}
  \nu = \frac{\sum_{v \in V}(d(v))^2}{d^2 n}.
\end{equation}
This can also be written as $\nu =\brac{n \;\sum_{v \in V}d^2(v)}/(2m)^2$.
The parameter $\nu$
 ranges from $1$ for regular graphs to $\Theta(n)$ for a star graph.
We prove the following  general theorem.
\begin{theorem}\label{Coalesce-expan}
Let $G$ be a connected graph with $n$ vertices, $m$ edges,
and let
$\nu =\brac{n \;\sum_{v \in V}d^2(v)}/(2m)^2$.
Let $C(n)$ be the expected coalescence time for a system of $n$ particles making a
lazy random walk on $G$, where originally one particle starts at each vertex.
Then
\beq{Ekval}
C(n) = O\brac{\frac{1}{1-\l_2}\brac{\log^{4} n +  \frac{n}{\nu}}}.
\eeq
By the equivalence between coalescence and voting, the expected time
$\E(C_{\ul v})$ to complete voting on $G$
has the same upper bound as $C(n)$.
\end{theorem}

Although Theorem \ref{Coalesce-expan} is a general statement of our results,
the bound  \eqref{Ekval} can be improved in extremal cases.
It is established in \eqref{altCn} of Section \ref{mainprf}
 that
\begin{equation}\label{ren2341}
C(n)= O\brac{\frac{1}{1-\l_2}\brac{\frac{m}{\D}\log n}^2}.
\end{equation}
This bound is better than~(\ref{Ekval}), if $\D = \omega(m/\log n)$.
For example, for a star,
(\ref{Ekval}) gives
$C(n)=O(\log^4 n)$ and
(\ref{ren2341}) gives
$C(n)=O(\log^2 n)$, whereas the correct value is $C(n)=\Th(\log n)$
(since a star is a bipartite graph, we consider the lazy walk).

Hassin and Peleg~\cite{HassinPeleg-InfComp2001} showed that voting (hence also coalescence)
is completed in expected $O(n^3 \log n)$ time on any connected graph.
The bound \eqref{Ekval} is parameterized by the eigenvalue gap, and can offer a refinement of Hassin and Peleg's
bound.
As $1-\l_2=\Omega( 1/n^2)$ for any
connected regular graph,
coalescence for these graphs is completed in $O(n^3)$ expected time.
 An example of a (non-regular) graph with coalescence time $\Th(n^3)$ is given by
two cliques of size $n/4$ joined by a path of length $n/2$. On the other hand
$1-\l_2=\Theta(1/n^3)$ for lollypop graphs, indicating that the bound \eqref{Ekval} is not tight.

The parameter $\nu$
is related to the second moment of
the degree distribution and measures the variability of the degree sequence.
If $\D$ is the  maximum degree of $G$, then $1 \le \nu \le \D/d \le n$.
For {\em near regular\/} graphs, when the ratio of the largest to the smallest vertex
degree is bounded by a constant, we have $\nu \le \D/d = O(1)$,
so the bound~\eqref{Ekval} becomes
\[
C(n) = O \bfrac{n}{1-\l_2}.
\]
In particular,
if $G$ is an expander in the
classic sense that it is regular and its eigenvalue gap $(1-\l_2)$ is constant, then $C(n)=O(n)$.

In parallel with our work, Oliveira~\cite{Oliveira-2012}  recently proved
the conjecture $C(n)=O(H_{\max})$  for  continu\-ous-time random walks.
The  result of
Oliveira implies an analogous linear bound $C(n)=O(n)$ for continuous-time random walks
on expanders.

We note that the bound \eqref{Ekval} is qualitatively different from $O(H_{\max})$, as
the graph structure is made explicit through the parameter $\nu$.
As $H_{\max}= \Om(n)$ for any graph (see~(\ref{nciwbc34})), the bound \eqref{Ekval}
 can improve on  $C(n)= O(H_{\max})$. This can occur for example if $\nu = \omega(1)$,
but also when $\nu = \Theta(1)$,
since there are graphs with $H_{\max} = \omega(n/(1-\l_2))$. Some examples follow.

For graphs with a power law (heavy tailed) degree distribution, Theorem~\ref{Coalesce-expan} can
give sublinear bounds on the coalescence and voting times as the following example shows.
Mihail {\it et al.}~\cite{Mihail-etal-2003} prove
that for $2 < \alpha < 3$, the random
$\Theta(n)$-vertex graph
with $\lceil n/d^{\alpha}\rceil$ vertices of degree $d$, for $d = 3,4, \ldots, n^{1/2}$,
has an $\Omega(\log^{-2} n)$ eigenvalue gap.
For this class of power law graphs,
$\nu = \Theta\brac{n^{(3-\alpha)/2}}$,
so Theorem~\ref{Coalesce-expan} implies a sublinear $O(n^{(\alpha-1)/2}\log^2 n)$
voting time, whereas for any graph, $H_{\max} = \Omega(n)$.

There are also examples of graphs with $\nu = \Theta(1)$ for which
our bound is asymptotically better than $O(H_{\max})$.
Consider the graph consisting of $(\log n)$-degree expander ($1-\l_2 \le c < 1$) with an additional vertex
attached to one of the vertices of the expander.
For this graph $\nu = \Theta(1)$ and $1-\l_2$ is a positive constant,
so $C(n) = O(n)$, but $H_{\max} = \Theta(n\log n)$.

It would be interesting to have a general lower bound on $C(n)$ which incorporates the graph structure
in a similar way to the upper bounds~\eqref{Ekval} and \eqref{ren2341}, but it is not clear what
form such a bound might take. A weaker conjecture is $C(n)= \Om(1/(1-\l_2))$.
This bound is tight for a path on $n$ vertices, where
$1/(1-\l_2) = \Th(n^2)$ and
$C(n)= C(2)= \Th(n^2)$.
Indeed
$C(n)\le n^2$, the cover time of the graph by a particle starting from the left most vertex;
and $C(2) \ge n^2/4$, the expected hitting time of the central vertex by  particles starting from the left most and right most vertices.

\subsection*{Structure of the paper}

The analysis of the coalescence process
(that is, the proof of Theorem~\ref{Coalesce-expan})
is divided into two phases.
During the first phase the number of particles decreases from the initial $n$ to a threshold value $k^*$.
This phase is analysed by showing that for a suitably chosen number of steps $t^* = t^*(k^*)$,
the probability that there exist
$k^*$ particles which do not have a single meeting between them within the first $t^*$ steps
is at most $1/2$.
This implies that with probability at least $1/2$, the number of particles at step $t^*$ is less than $k^*$.

The second phase, when the number of particles decreases from $k^*$ to $1$, is analysed by
bounding the expected time we have to wait until the first meeting between any of $k$ particles,
where $2 \le k \le k^*$. At the time of this first meeting,
the number of particles decreases from $k$ to $k-1$
(with some relatively small probability, the first meeting could involve more than $2$ particles,
reducing the number of particles to fewer than $k-1$).
The analysis of the second phase is
based on the following theorem
bounding the expected time to first meeting between any of $k$ particles.

\begin{theorem}\label{General}
Let $k^*$ be given by
\begin{equation}\label{bqhbf}
k^* = \max\left\{ 2, \;
                                \min\left\{ \bfrac{n}{\nu}^{1/2},
                                                 \; \frac{m}{2\D},
                                                 \; \log n \right\}\right\},
\end{equation}
where $\D$ is the maximum degree,
and $\nu$ given by~(\ref{nu-definition}).
For $2 \le k \le k^*$ particles starting from arbitrary vertices in $G$,
let $M_k$ be the time to first meeting.
Then
\begin{equation}\label{EMK}
\E (M_k) =O\brac{\frac{1}{1-\l_2}\brac{ {k \log n}+ \frac{n}{\nu k^2}}}.
\end{equation}
\end{theorem}

The expression~(\ref{bqhbf}) for the threshold value $k^*$ is not very transparent,
but seems to be necessary to deal with the
generality of degree sequences of connected graphs.
Provided the maximum degree of the graph satisfies $\D \le 2m/\log^2 n$,
then $k^*=\log n$. The other terms are there to cover extremal cases such as star graphs. The condition that $k^* \ge 2$ ensures
 there are at least 2 particles to coalesce.

Section \ref{visit} gives background material on random walks. Section \ref{multiwalk}
replaces multiple random walks by a single walk on a suitably defined product graph.
Theorem~\ref{General} is proven in Section \ref{rets}
and the proof of Theorem~\ref{Coalesce-expan} is concluded in Section~\ref{mainprf}.

\section{Random walk properties}\label{visit}

Let $G=(V,E)$ denote a connected undirected graph, $|V|=n$, $|E|=m$, and
let $d(v)$ be the degree of a vertex $v$.
A {\em simple random walk} $\cW_u,\,u\in V$, on
graph $G$ is a Markov chain
modeled by a particle moving from vertex to vertex according
to the following rule.
The probability of transition from  vertex $v$ to  vertex $w$
is equal to $1/d(v)$, if $w$ is a neighbour of $v$, and
$0$ otherwise.
The walk $\cW_u$
starts from vertex $u$ at $t=0$.
Denote by $\cW_u(t)$ the vertex reached at step $t$;
$\cW_u(0)=u$.

We assume $G$ is connected, and the random walk $\cW_{u}$ on $G$ is ergodic with stationary distribution $\pi$,
where $\pi_v=d(v)/(2m)$. If this is not the case, e.g. $G$ is  bipartite, then the walk can be made ergodic,
by making it lazy.
A  random walk is {\em lazy}, if it moves from $v$ to one of its neighbours $w$ with probability $1/(2d(v))$,
and stays at vertex $v$ with probability $1/2$.

Let $P=P(G)$ be the matrix of
transition probabilities of the walk and let
$P_{u}^{t}(v)=\Pr(\cW_{u}(t)=v)$.
The eigenvalues of $P(G)$ are real, and can be ordered $\l_1=1 >
 \l_2\ge\cdots\ge \l_n$, where $ \l_{n} > -1$ as the walk is ergodic.
 Let
$\l=\max(\l_2,|\l_{n}|) < 1$.
The rate of convergence of the walk is given by
\begin{equation}\label{mix}
|P_{u}^{t}(x)-\pi_x| \leq (\p_x/\p_u)^{1/2}\l^t,
\end{equation}
where $|r|$ is the absolute value of the real number $r$.
For a proof
see for example,  Lovasz \cite{Lo} Theorem 5.1.
We assume henceforth that $\l=\l_2$. If not,
 the standard way to ensure that  $\l=\l_2= \l_2(G)$, is to make the chain  lazy.

We use the following definition of mixing time $T_G$, for a graph $G$.
For all vertices $u$ and $x$ in $G$ and any $t \ge T_G$,
\begin{equation}\label{mixing}
  |{P}_{u}^{(t)}(x)-{\pi}_x| \leq o\bfrac{1}{n^2}.
\end{equation}
For convenience we assume that $T_G= \Om( \log n)$, even if this is
not necessary.

Let $\E_{\pi}(H_w)$ denote the expected hitting time of a vertex $w$ from
the stationary distribution $\pi$.
The quantity $\E_{\pi}(H_w)$
 can be expressed as (see e.g.~\cite{AlFi}, Chapter 2)
\begin{equation}\label{pi-hit}
\E_{\pi}(H_v) = Z_{vv}/\pi_v,
\end{equation}
where
\begin{equation} \label{pihitz}
Z_{vv} = \sum_{t=0}^{\infty} (P_{v}^{(t)}(v) - \pi_v).
\end{equation}

Let $A_v(t;u)$ denote the event that $\cW_u$ does not visit vertex $v$ in steps $0,...,t$.
The following lemma gives a bound
on the probability of this event
in terms of $\E_{\pi}(H_v)$
and the mixing time of the walk.
\begin{lemma}\label{crude}
Let $T = T_G$ be a mixing time of a random walk $\cW_u$ on $G$ satisfying \eqref{mixing}.
Then
\[
\Pr(A_v(t;u)) \; \le \; e^{-\rdown{{t}/{(T+3\E_{\pi}(H_v))}}}.
\]
\end{lemma}

\proofstart
Let $\r \equiv P_u^{(T)}$ be the distribution of $\cW_u$ on $G$ after $T$ steps.
Then~\eqref{mixing} and the fact that
$\pi_x \ge 1/n^2$
for any connected graph imply
\beq{this-eqn}
\E_{\r}(H_v)=\ooi \E_{\pi}(H_v).
\eeq

Let $H_v(\r)$ be the time to hit $v$ starting from $\r$,
and let $\t=T+3\E_{\pi}(H_v)$. Then, noting that $\E_{\r}(H_v) \equiv \E(H_v(\r))$,
\begin{eqnarray*}
\Pr(A_v(\t;u)) & = & \Pr(\,A_v(T;u) \; \mbox{and} \; H_v(\r) \ge 3\E_{\pi}(H_v)\,) \\
   & \le & \Pr\brac{\,H_v(\r) \ge 3 \E_{\pi}(H_v)\,} \\
   & \le & \Pr\brac{\,H_v(\r) \ge e \cdot \E(H_v(\r))\,} \\
   & \le & \frac{1 }{e}. \\
\end{eqnarray*}
By restarting the process  $\cW_u$ at
$\cW_u(0)=u$, $\cW_u(\t)$, $\cW_u(2\t)$, $\ldots, \cW_u((\rdown{t/\t}-1)\t)$, we obtain
\[
\Pr(A_v(t;u)) \le  e^{-\rdown{t/\t}}.
\]
\proofend

\section{Multiple random walks}\label{multiwalk}

We consider the coalescence of $k\ge 2$ independent random
walks on a graph $G=(V_G,E_G)$. To do this we replace the $k$ walks by a single walk
as follows.

Let graph $Q = Q_k=(V_Q,E_Q)$
have vertex set $V_Q=V^k$.
Thus a vertex $\ul v$ of $Q_k$ is a $k$-tuple
$\ul v= (v_1,v_2,...,v_k)$ of vertices $v_i \in V_G, i=1,...,k$,
with repeats allowed.
Two vertices $\ul v, \ul w \in V_Q$ are adjacent if  $\{v_1,w_1\},
...,\{v_k,w_k\}$ are edges of $G$.
There is a direct equivalence between
$k$ random walks  $\cW_{u_i}(t)$ on $G$ with
starting positions  $u_i$
and a single random walk $\cW_{\sul u}(t)$ on $Q_k$ with starting position
$\ul u= (u_1,u_2,...,u_k)$.

For any starting positions $\ul u=(u_1,...,u_k)$ of the walks,
let $M_k(\ul u)$ be the time until the first meeting in $G$.
Let $S_k \subseteq V(Q_k)$, the {\em diagonal set of vertices}, be defined by
\[
S = S_k=\{(v_1,...,v_k): v_i=v_j \text{ some } 1\le i< j \le k\}.
\]
If the random walk on $Q_k$ visits this set, two particles occupy the same vertex in the underlying
graph $G$ and a (coalescing) meeting occurs.

The number of  visits to the set $S_k$  by a random walk is not a readily manipulated quantity.
An easier approach is to contract $S_k$ to a single vertex
$\g = \g_k = \g(S_k)$,  thus replacing $Q_k$
by a graph $\G = \G_k$.
On contraction, all edges, including loops, are retained.
Thus $d_{\G}(\g)=d_Q(S)$, where $d_F$ denotes vertex degree in graph $F$,
and the degree $d_F(X)$ of a set $X$ is the sum of the degrees of the vertices
in $X$.
Moreover $\G$ and $Q$ have the same total degree, and
the degree of any vertex of $\G$ other than $\g$ is the same as in graph $Q$.
Let $\pi$ and $\hat{\pi}$ be the stationary distributions of a random walk on $Q$
and $\G$, respectively.
If ${\ul v} \not \in S$
then $\hat \pi_{\sul v} = \pi_{\sul v}$, and $\hat \pi_{\g}=\pi_S \equiv \sum_{{\sul x}\in S} \pi_{\sul x}$.

It follows that, if $T_\G$ is a mixing time satisfying~\eqref{mixing} in $\G$, then
\beq{EMk}
\E (M_k(\ul u)) \le T_{\G}+ \ooi \E_{\wh \pi} (H_{\g_k}),
\eeq
where $\E_{\wh \pi} (H_{\g_k})$ is the hitting time of $\g_k$ in $\G$ from stationarity.

Since we have replaced $k$ individual walks on $G$  by a single walk
on $Q_k$, and then on $\G$, we need to
relate mixing times on $T_Q$ and  $T_{\G}$ directly to
 a given mixing time
$T_G$ of a single random walk on the underlying graph~$G$.
(We will need $T_{\G}$ in two places:
in the bound~(\ref{EMk})
and when applying Lemma~\ref{crude} to graph $\G$.)

\begin{lemma}\label{T-value}
For random walks in graphs $G$, $Q$ and $\G$,
there are mixing times
\begin{equation}\label{mixtimeQGamma}
 T_G = O\brac{\frac{\log n}{1-\l_2(G)}},\: T_Q=O(kT_G),\: T_{\G}=O(k T_G),
\end{equation}
such that
\[
\max_{u,x\in V_F}|P_{u}^{t}(x)-\pi_x| =o(1/n_F^{2}),\;\;\; \mbox{for any $t\geq T_F$},
\]
where $F$ is any of the graphs $G$, $Q$ or $\G$, and $n_F = |V_F|$.
\end{lemma}

\proofstart
The bound on $T_G$ is well known (see for example, Sinclair~\cite{Sinclair}):
use~\eqref{mix}, observing that $\pi_x/\pi_u = O(n)$
and $\l_2^{1/(1-\l_2)}$ has a constant $c < 1$ upper bound.
To use \eqref{mix} also to
derive bounds on $T_Q$ and $T_\G$,
we need to know the eigenvalues of $Q_k$ and $\G$ in terms of the eigenvalues of $G$.
We have $\l_2(\G) \le \l_2(Q_k)$ and $\l_2(Q_k) = \l_2(G)$.
This follows from established results, as we next explain.

In the notation of Markov processes, the random walk on $Q_k$ is known as the
{\em tensor product chain},
and its eigenvalues are the $k$-wise products of the eigenvalues of $G$. Thus, assuming $\l_2(G)\ge\l_n(G)$,
it follows that $\l_2(Q_k)=\l_2(G)$. See \cite{LPW} page 168 for more details.

In the notation of \cite[Ch.\ 3]{AlFi},
the random walk on $\G$ is the random walk on $Q_k$ with $S$ collapsed to $\g(S)$.
It is proved in \cite[Ch.\ 3]{AlFi}, Corollary 27,
that if a subset $A$ of vertices is collapsed to a single vertex,
then the second eigenvalue of the transition matrix cannot increase
(in that corollary the variable $\t_2=1/(1-\l_2)$).
Thus $\l_2(Q) \ge \l_2(\G)$.

We get the factor $k$ in the bounds~\eqref{mixtimeQGamma} on the mixing times
$T_Q$ and $T_\G$,
because $\pi_x/\pi_u = O(n^{2k})$ and we need $|P_{u}^{T}(x)-\pi_x| = o(1/n^{2k})$, as
the number of vertices in graphs $Q$ and $\G$ is $O(n^k)$.
\proofend

\vspace{0.15in}

For reference, we record the salient facts for the graphs $G, Q,\G$ in Table~\ref{table}.
The bound on $\pi_\g$
will be established in Lemma \ref{dgamma}.

\begin{table*}
\begin{center}
\begin{tabular}{|l||l|l|l|} \hline
Graph & vertices & Stationary distribution  $\pi$ & Mixing time \\
\hline
$G$&$n_G=n$&$\pi_v=d(v)/2m$& $T_G=O(\log n/(1-\l_2))$\\
$Q_k$&$n_Q=n^k$& $\pi_{\sul v} = d(v_1)\cdots d(v_k)/(2m)^k$&$T_Q=O(k T_G)$\\
$\G_k$&$n_\G \le n^k$& $\pi_\g \ge k^2\nu/(8n)$&$T_\G=O(T_Q)$\\
\hline
\end{tabular}
\end{center}
\caption{The main parameters of the random walks on graphs $G$, $Q_k$ and $\G_k$.\label{table}}
\end{table*}

\section{Hitting time from stationarity -- \, Proof of Theorem~2}\label{rets}

The proof of Theorem~\ref{General}
is based on Inequality~(\ref{EMk})
and on a good upper bound on the expected hitting time of vertex $\g$ by
a random walk in $\G$ which starts from the stationary distribution.
We obtain such a bound using~\eqref{pi-hit} by deriving an upper
bound on $Z_{\g\g}$ (Lemma~\ref{Eval-gap}) and
a lower bound on the stationary probability $\pi_\g = \wh\pi_\g$ (Lemma~\ref{dgamma}).

\begin{lemma}\label{Eval-gap}
Let $F$ be a graph with the eigenvalue gap $1-\l_2$,
 then
\begin{equation}\label{hitit-eval}
Z_{vv} \le  \frac{1}{1-\l_2}.
\end{equation}
In particular, for any vertex $v$ of $G,$ $Q$ or $\G$,
$Z_{vv} \le  1/(1-\l_2(G))$.
\end{lemma}

\proofstart
Let $\l_2=\l_2(F)$.
Using \eqref{mix} with $x=u=v$ gives
\[
|P_v^t(v)-\pi_v| \le \l_2^t,
\]
and thus
\[
Z_{vv}=\sum_{t \ge 0} (P_v^t(v)-\pi_v) \le \sum_{t \ge 0} \l_2^t = \frac{1}{1-\l_2}.
\]
The proof of Lemma \ref{T-value} establishes that
$(1-\l_2(\G)) \ge 1-\l_2(Q) = 1-\l_2(G)$.
\proofend

\begin{lemma}\label{dgamma}
Let $G$ be a connected graph with $n$ vertices and $m$ edges.
Let
\beq{kstar}
k^* = \max\left\{ 2, \;
                                \min\left\{ \bfrac{n}{\nu}^{1/2},
                                                 \; \frac{m}{2\D},
                                                 \; \log n \right\}\right\},
\eeq
where $\D$ is the maximum degree of $G$ and
$\nu= (n/(2m)^2)\sum_{v\in V} d^2(v)$.
Let $k$ be integer, $2 \le k \le k^*$.
Let $\g = \g_k$ in $\G$
be the contraction of $S = S_k$ in $Q$. Then  
\begin{equation}\label{stationarybound}
\pi_\g \; = \; \frac{d(\g)}{(2m)^k}
  \; \ge \; \frac{k^2\nu}{8n}.
\end{equation}
\end{lemma}

\proofstart
By definition, $d(\g)=d(S)$.
If $k = 2$, then,
\[
d(S) =  \sum_{v\in V} d^2(v) = (2m)^2\frac{\nu}{n}.
\]
If $3 \le k \le k^*$,
for $1 \le x < y \le k$, define the following subsets of $S$:
\[
S_{(x,y)} \; = \;  \{ (v_1,\ldots, v_k): \: v_x = v_y \}.
\]
We have
\[ S = \bigcup_{1 \le x < y \le k} S_{(x,y)},
\]
and
\[
d\brac{S_{(x,y)}} \; =  \; (2m)^{k-2}\sum_{v\in V} d^2(v)= (2m)^k\frac{\nu}{n}.
\]
For $\{x,y\} \neq \{p,q\}$, $d\brac{S_{(x,y)} \cap S_{(p,q)}}$ equals to
\[
\begin{array}{ll}
   (2m)^{k-4} \sum_{v,u\in V} d^2(v) d^2(u), & \mbox{if $\{x,y\}\cap\{p,q\}=\emptyset$, or} \\
   (2m)^{k-3} \sum_{v\in V} d^3(v), & \mbox{if $|\{x,y\}\cap\{p,q\}|=1$.} \rule[-1ex]{0em}{4ex}
          \end{array}
\]
Therefore, from the inclusion-exclusion principle,
\begin{align}
 d(S) & \ge  \sum_{\{x,y\}} d\brac{S_{(x,y)}}
        - \sum_{\{x,y\}\neq\{p,q\}} d\brac{S_{(x,y)} \cap S_{(p,q)}} \nonumber \\
     & \ge  {k\choose 2} (2m)^{k} \frac{\nu}{n}
       - 3 {k\choose 4} (2m)^{k} \frac{\nu^2}{n^2}
      - \, 3 {k\choose 3} (2m)^{k} \frac{\D\nu}{2mn} \label{why3} \\
     & \ge  {k\choose 2} (2m)^{k} \frac{\nu}{n}
         \brac{1 - \frac{k^2\nu}{4n} - \frac{k\D}{2m}} \label{lbFordS} \\
     & \ge  {k\choose 2} (2m)^{k} \frac{\nu}{2n}.
                  \label{degreebound}
\end{align}
The factor 3 in \eqref{why3} occurs as the number of ways to partition 4 objects into  disjoint sets of size 2, and
partition 3 objects into sets of size 2 with single intersection, respectively.
The bound~\eqref{degreebound} follows from~\eqref{lbFordS}, by noting the upper bound on $k$ in \eqref{kstar}.
\proofend

{\bf Proof of Theorem~\ref{General}.}
Let $M_k$ be the time of the first meeting
among $k\le k^*$ particles in $G$,  and
let $\g = \g_k$ be the contraction of the diagonal set $S=S_k$.
Using \eqref{pi-hit} for graph $\G$ and with $v = \g$, and Lemmas~\ref{Eval-gap}
and~\ref{dgamma}
we have, that the  hitting time $H_{\g}$ of $\g$ from stationarity has expected value
\begin{eqnarray}
\E_{\pi} (H_{\g}) & \le & \frac{1}{\pi(\g)} \, \frac{1}{1-\l_2} \label{piggy} \\
    & \le & \frac{8}{k^2}\, \frac{n}{\nu} \, \frac{1}{1-\l_2}. \label{expc-1st-meet}
\end{eqnarray}
Since $T_{\G}= O(kT_G)$, and referring to \eqref{EMk} and Table \ref{table},
\begin{eqnarray}
\E (M_k) & \leq & O(k T_{G}) +\ooi \E_{\pi} (H_\g) \label{poggy} \\
  & = & O\brac{\frac{1}{1-\l_2}\brac{ {k \log n}+\frac{n}{\nu k^2}}}. \label{expectedmeet}
\end{eqnarray}
\proofend

Let $C_k$ be the time for $k \le k^*$ particles to coalesce.
For use in the proof of Theorem~1 in the next section, we state an upper bound on $\E(C_{k})$
which follows directly from Theorem~\ref{General}.
Using~\eqref{expectedmeet}
and noting that $\sum_s (1/s^2)\le \pi^2/6$ is constant, we have
\begin{equation}\label{expecCk}
\E(C_{k}) \;\; \le \;\; \sum_{s=2}^{k} \E (M_s)
\;\; = \;\; O\brac{\frac{1}{1-\l_2}\brac{ {k^2 \log n}+\frac{n}{\nu}}}.
\end{equation}

\section{Coalescence time: Proof of Theorem~1}\label{mainprf}

We consider the case of $n$ coalescing particles, where each particle is
initially located at a distinct vertex of the graph.
The purpose of this section is to conclude the proof that for any connected graph
\beq{Ekval2}
C(n) = O\brac{\frac{1}{1-\l_2}\brac{\log^4 n +  \frac{n}{\nu}}}.
\eeq
To establish this result, we first prove that the probability
that there exist
$k^*$ particles which do not have a single meeting between them within the first $t^*$ steps
is at most $1/2$, if
\[
t^* =  k^*\log n\brac{ T_\G+ 3\E_\pi(H_{\g})},
\]
where $\G = \G_{k^*}$, $\g = \g_{k^*}$ and
the value of $k^*$ is given in \eqref{kstar}.
An upper bound on the expected time  $\E (C_{k})$ for $k \le  k^*$ particles to coalesce is
 given in \eqref{expecCk} above, and we can deal
with that part separately.

Let ${\cal P} = {\cal P}(\ul v)$
be the set of $k^*$ particles
starting from vertices $\ul v=(v_1,...,v_{k^*})$.
The probability that the particles in ${\cal P}$ do not meet by time $t$ is the same as
the probability that the random walk in $\G$ starting from $\ul v$ does not
visit $\g$ by time $t$.
We apply Lemma~\ref{crude} to graph $\G$, vertex $\g$ and $t = t^*$,
and obtained that
\begin{eqnarray*}
\lefteqn{\Pr(\text{no meeting among particles in } {\cal P} \text{ before } t^*)}
\;\;\;\;\;\;\;\;\;\;\;\;\;\;\;\;\;\;\;\;\;\;\;\; \quad\quad\quad\quad\quad\quad\quad\quad \\
& \le &  e^{-k^*\log n} \; = \; n^{-k^*}.
\end{eqnarray*}

In the coalescence process, we can assume that if two or more particles meet at the same vertex,
then the lowest index particle survives (and continues its random walk) while the other particles die.
Thus if there are $k^*$ or more particles after $t^*$ steps, then there is a set ${\cal P}$ of $k^*$
particles which do not meet within $t^*$ steps.
Therefore,
\begin{eqnarray}
\lefteqn{\Pr(\text{at least $k^*$ particles after $t^*$ steps})} \nonumber \\
& \le & \Pr(\text{exists a set $\cal P$ of $k^*$ particles
with no meeting before $t^*$})  \nonumber\\
& \le &  {n \choose k^*} n^{-k^*} \; \le \; \frac{1}{2}. \label{nvjevev7}
\end{eqnarray}
The last inequality holds  because ${n \choose k} \le n^k/k!$ and $k^* \ge 2$.
The bound~(\ref{nvjevev7})
implies that the expected number of steps until fewer than $k^*$ particles remain
is at most $t^* + \frac{1}{2}(2t^*) +  \frac{1}{4}(3t^*) + \cdots = 4t^*$.
Therefore, using $T_\G = O(k^* \log n/(1-\l_2))$ from Lemma \ref{T-value},
the bound on $\E_\pi(H_\g)$ given in~(\ref{expc-1st-meet}),
and the bound on $\E(C_{k^*})$ given in~(\ref{expecCk}),
we obtain the bound~(\ref{Ekval2}):
\begin{eqnarray}
C(n) & \le & 4t^* + \E(C_{k^*}) \nonumber \\
        & = & O\brac{\frac{(k^*\log n)^2}{1-\l_2}
                                + \frac{1}{1-\l_2}\frac{\log n}{k^*}\frac{n}{\nu}
                                + \frac{1}{1-\l_2}\brac{(k^*)^2\log n + \frac{n}{\nu}}} \nonumber \\
       & = & O\brac{\frac{1}{1-\l_2}\brac{(k^* \log n)^2
                              + \frac{\log n}{k^*}\frac{n}{\nu}}} \label{ghsw} \\
       & = & O\brac{\frac{1}{1-\l_2}\brac{\log^4 n + \frac{n}{\nu}}}. \nonumber
\end{eqnarray}
The last bound above is obvious if $k^* = \log n$.
If  $k^* < \log n$, then
the last bound holds because
the second term in the sum in~(\ref{ghsw}), that is $(n/\nu)\log n/k^*$, is $O(\log^3 n)$.
Indeed, if $k^* < \log n$, then from the definition of $k^*$, either
$\bfrac{n}{\nu}^{1/2} < \log n$ or
$\frac{m}{2\D} < \log n$.
If the former, then the second term in the sum in~(\ref{ghsw}) is clearly $O(\log^3 n)$.
Observe that
\begin{equation}\label{njjnfv674}
    \frac{n}{\nu} \le \frac{n^2 d^2}{\D^2} = \bfrac{2m}{\D}^2.
\end{equation}
Thus if  $\frac{m}{2\D} < \log n$, then
$n/\nu = O(\log^2 n)$, and
the second term in the sum in~(\ref{ghsw}) is again $O(\log^3 n)$.

We conclude by noting that since $k^* \le \frac{m}{2\D}$ and (\ref{njjnfv674}),
then~(\ref{ghsw}) implies
\beq{altCn}
C(n)= O\brac{\frac{1}{1-\l_2}\brac{\frac{m}{\D}\log n}^2}.
\eeq
The above bound is better than~(\ref{Ekval2}), if $\D = \omega(m/\log n)$.



\begin{thebibliography}{10}


\bibitem{Aldous-MeetingTimes} D. Aldous.
Meeting times for independent Markov chains.
{\em Stochastic Processes and their Applications\/} 38(2):185-193, August 1991.

%
\bibitem{AlFi} D. Aldous and J. Fill. {\em Reversible Markov Chains and
Random Walks on Graphs}, \\
{\tt http://stat-www.berkeley.edu/pub/users/aldous/RWG/book.html.}



\bibitem{CFR}  C. Cooper, A. M. Frieze, and T. Radzik.
Multiple Random Walks in Random Regular Graphs.
{\em SIAM J. Discrete Math.} 23(4):1738-1761, 2009.
%

\bibitem{CEOR-PODC2012} C. Cooper, R. Els\"asser, H. Ono, T. Radzik.
Coalescing random walks and voting on graphs.
In {\em  PODC 2012: Proceedings of the 2012 ACM
Symposium on Principles of Distributed Computing}, pages 47-56, July 2012.

\bibitem{Cox-1989} J. T. Cox.
Coalescing random walks and voter model consensus times on the torus
in $\mathbb{Z}^d$.
{\em The Annals of Probability\/} 17(4):1333-1366, October 1989.


\ignore{
\bibitem{DonnellyWelsh-1983} P. Donnelly and D. Welsh.
Finite particle systems and infection models.
{\em Math. Proc. Camb. Phil. Soc.}  94(1):167-182, July 1983.
}

\bibitem{Mihail-etal-2003}
C.~Gkantsidis, M.~Mihail, and A.~Saberi.
Conductance and congestion in power law graphs.
In {\em SIGMETRICS 2003: Proceedings of \ 2003 ACM SIGMETRICS Intl.\ Conf.\
on Measurement and Modeling of Computer Systems,} New York, NY, USA,
pages 148-159, 2003.


\bibitem{HassinPeleg-InfComp2001}
Y.~ Hassin and D.~Peleg.
Distributed probabilistic polling and applications to proportionate agreement.
{\em Information \& Computation} 171(2):248-268, December 2001.



\bibitem{IJ} A. Israeli and M. Jalfon. {Token management schemes and
random walks yeild self stabilizing
mutual exclusion.} In {\em PODC 1990: Proceedings of the 9th Annual ACM
Symposium on Principles of Distributed Computing}, Quebec City,
Quebec, Canada, pages 119-131, August 1990.



\bibitem{LPW} D. Levin, Y. Peres, and E. Wilmer. {\em Markov Chains
and Mixing Times.}
American Mathematical Society, 2009.

\bibitem{Lo} L. Lov\'asz. Random walks on graphs: a survey.
{\em Bolyai Society Mathematical Studies.}
{Combinatorics,  Paul Erd\H{o}s is Eighty} 2:1-46, Keszthely, Hungary, 1993.

\ignore{
\bibitem{Nakata_etal_1999}
T. Nakata, H. Imahayashi, M. Yamashita.
Probabilistic local majority voting for the agreement problem on
finite graphs.
In {\em COCOON 1999: Proceedings of the 5th Annual International
Conference on Computing and Combinatorics}, Tokyo, Japan, pages
330-338, July 1999.
}

\bibitem{Oliveira-2012}
R.~Oliveira.
On the coalescence time of reversible random walks.
{\em Trans. Amer. Math. Soc.\/} 364(4): 2109-2128, 2012.



\bibitem{Sinclair} A. Sinclair. Improved bounds
for mixing rates of Markov chains and multicommodity flow.
{\em Combinatorics, Probability and Computing} 1(4):351-370, December 1992.


\end{thebibliography}
\end{document}